\documentclass[superscriptaddress,twocolumn,showpacs,a4paper,
amssymb,amsmath,nobibnotes,aps,prd,
showkeys,
nofootinbib,notitlepage]{revtex4-1}
\usepackage{verbatim}
\usepackage[T1]{fontenc}
\usepackage[utf8]{inputenc}
\usepackage[american]{babel}
\usepackage{epsfig}
\usepackage{graphicx,subcaption,caption}
\usepackage{booktabs}
\usepackage{multirow}
\usepackage{dcolumn}
\usepackage{amsmath}
\usepackage{mathtools}
\usepackage{amsfonts}
\usepackage{amssymb}
\usepackage{epstopdf}
\usepackage{bm}
\usepackage{siunitx}
\usepackage{braket}
\usepackage{enumitem}
\usepackage{soul}
\usepackage{color}

\ExplSyntaxOn 
\cs_new:Npn \expandableinput #1
  { \use:c { @@input } { \file_full_name:n {#1} } }
\ExplSyntaxOff

\makeatletter 
\providecommand*{\input@path}{}
\g@addto@macro\input@path{{"C:/Users/me/inpath/"}}
\makeatother

\definecolor{navyblue}{rgb}{0.0, 0.0, 0.5}
\definecolor{royalblue}{rgb}{0.25, 0.41, 0.88}
\definecolor{cadmiumgreen}{rgb}{0.0, 0.42, 0.24}
\definecolor{blue-violet}{rgb}{0.54, 0.17, 0.89}
\definecolor{darkviolet}{rgb}{0.58, 0.0, 0.83}
\definecolor{orange(colorwheel)}{rgb}{1.0, 0.5, 0.0}

\usepackage{hyperref}
\hypersetup{
    colorlinks=true, 
    linkcolor=royalblue, 
    citecolor=magenta}

\usepackage{booktabs}
\usepackage{multirow}
\usepackage{dcolumn}
\usepackage{colortbl}

\newcommand{\mnu}{\ensuremath{\Sigma m_{\nu}}}
\newcommand{\neff}{\ensuremath{N_{\rm eff}}}

\begin{document}

\title{Impact of the damping tail on neutrino mass constraints}

\date{\today}

\author{Eleonora Di Valentino}
\email{e.divalentino@sheffield.ac.uk}
\affiliation{School of Mathematics and Statistics, University of Sheffield, Hounsfield Road, Sheffield S3 7RH, United Kingdom}

\author{Stefano Gariazzo}
\email{gariazzo@to.infn.it}
\affiliation{Istituto Nazionale di Fisica Nucleare (INFN), Sezione di Torino, Via P.\ Giuria 1, I-10125 Turin, Italy}

\author{William Giar\`e}
\email{w.giare@sheffield.ac.uk}
\affiliation{School of Mathematics and Statistics, University of Sheffield, Hounsfield Road, Sheffield S3 7RH, United Kingdom}

\author{Olga Mena}
\email{omena@ific.uv.es}
\affiliation{Instituto de F{\'\i}sica Corpuscular  (CSIC-Universitat de Val{\`e}ncia), E-46980 Paterna, Spain}

\preprint{}

\begin{abstract}
Model-independent mass limits assess the robustness of current cosmological measurements of the neutrino mass scale. Consistency between high-multipole and low-multiple Cosmic Microwave Background observations measuring such scale further valuate the constraining power of present data. We derive here up-to-date limits on neutrino masses and abundances exploiting either the Data Release 4 of the Atacama Cosmology Telescope (ACT) or the South Pole Telescope polarization measurements from SPT-3G, envisaging different non-minimal background cosmologies and marginalizing over them. By combining these high-$\ell$ observations with Supernova Ia, Baryon Acoustic Oscillations (BAO), Redshift Space Distortions (RSD) and a prior on the reionization optical depth from WMAP data, we find that the marginalized bounds are competitive with those from Planck analyses. We obtain $\sum m_\nu <0.139$~eV and $N_{\textrm{eff}}= 2.82\pm 0.25$ in a dark energy quintessence scenario, both at $95\%$~CL. These limits translate into $\sum m_\nu <0.20$~eV and $N_{\textrm{eff}}= 2.79^{+0.30}_{-0.28}$ after marginalizing over a plethora of well-motivated fiducial models. Our findings reassess both the strength and the reliability of cosmological neutrino mass constraints. 
 \end{abstract}


\maketitle

\section{Introduction}
\label{sec:introduction}
Cosmological bounds on neutrino masses are reaching limits close to the lower bounds derived from neutrino oscillation
data~\cite{deSalas:2020pgw,Esteban:2020cvm,Capozzi:2021fjo}:
\begin{equation}\label{eq:lower_bound}
    \begin{split}
        \sum m_\nu > \left\{
        \begin{array}{ll}
(0.0591 \pm 0.00027) \, {\rm eV} &\quad \rm (NO) \\
(0.0997 \pm 0.00051) \, {\rm eV} &\quad \rm (IO) \\
        \end{array}\right.
    \end{split}\,.
\end{equation}
\newline
obtained by assuming that the lightest neutrino
mass is zero, and 

\begin{equation}\label{eq:oscillations}
\begin{split}
\Delta m^2_{21} &=(7.50\pm 0.21)\times 10^{-5} \,{\rm eV}^2 \,, \\
|\Delta m^2_{31}|&= \left\{
\begin{array}{ll}
(2.550\pm0.025)\times 10^{-3}\, {\rm eV}^2 &\quad \rm (NO) \\
(2.450\pm0.025)\times 10^{-3}\, {\rm eV}^2 &\quad \rm (IO)
\end{array}
\right. \,,
\end{split}    
\end{equation}
where $\Delta m^2_{ij} \equiv m_i^2-m_j^2$. The sign of $\Delta m^2_{31}$ determines the type of neutrino mass ordering,  being positive for normal ordering (NO) and negative for inverted ordering (IO). Interestingly, the tightest bound on $\sum m_\nu<0.09$~eV at $95\%$~CL~\cite{DiValentino:2021hoh,Palanque-Delabrouille:2019iyz,diValentino:2022njd} is comparable to the IO lower bound. In addition, one expects near future observations from ongoing galaxy surveys, such as DESI~\cite{DESI:2023bgx,DESI:2022lza}, to improve current limits, eventually reaching the NO predictions (in the absence of a signal). 
Even if data are not informative enough to claim a tension between cosmological and terrestrial, neutrino oscillation bounds~\cite{Gariazzo:2023joe}, it is timely to reassess the robustness of neutrino mass limits, as such a tension could strongly depend on the underlying fiducial cosmology.  In this regard, 
Bayesian model comparison techniques offer the ideal tool for computing model-marginalized cosmological parameter limits,
avoiding the biases due to the fiducial cosmology, see Ref.~\cite{Gariazzo:2018meg} for a pioneer study applied to neutrino mass limits. The  former method was extended to scenarios including  a freely varying neutrino mass abundances (parameterized via $N_{\rm eff}$) or a hot-dark matter axion component in Refs.~\cite{diValentino:2022njd}
and~\cite{DiValentino:2022edq}. However, in these previous studies the neutrino mass limits were driven by the input from Planck CMB data.  
 
In order to reinforce and, to some extent, to convey with particle physics constraints on the neutrino properties (mass, hierarchy and abundances), not only model-independent mass limits are required:  consistency tests between limits from high-multipole and low-multiple CMB probes are also mandatory. 
In this spirit, we present here model-marginalized limits on the neutrino mass $\sum m_\nu$ and on the relativistic degrees of freedom $N_{\rm eff}$ exploiting data from the ACT and SPT damping tail CMB experiments, analyzed in combination with the previous WMAP CMB probe. Other low redshift probes, such as Type Ia Supernovae and BAO will also be considered in the numerical analyses. The structure of the paper is as follows.
We start from \autoref{sec:bayesian}, which contains a dedicated description of the statistical method exploited here to derive model-marginalized bounds on the neutrino properties.
The following \autoref{sec:data} and \autoref{sec:models} describe the cosmological observations and fiducial cosmologies considered here, respectively,
while \autoref{sec:implementation} is devoted to our numerical implementation.
In \autoref{sec:results}
we illustrate and discuss the main results of our analyses, to conclude in \autoref{sec:conclusions}.

\section{Bayesian method}
\label{sec:bayesian}
In this study, we want to compare
how different cosmological models fit well different CMB data sets
and determine for each data combination a robust, model-marginalized constraint on some neutrino properties.
In order to do this, we follow the approach described for example by~\cite{Gariazzo:2018meg},
and summarized in the following.
Let us consider a set of models $\mathcal{M}_i$, which in our case are extensions of some initial model.
We will consider that all the models have the same prior probability.
Applying Bayesian model comparison,
the posterior probability for each model within the considered set can be written as
\begin{equation}
\label{eq:modelposterior}
p_i = \frac{Z_i}{\sum _j Z_j}
= \frac{B_{i0}}{\sum _j B_{j0}}
\,,
\end{equation}
where $Z_i$ is the Bayesian evidence of the $i$-th model
and $B_{i0}=Z_i/Z_0$ is the Bayes factor of model $i$ with respect to the favored model $\mathcal{M}_0$ within the set.
Notice that the sum on $j$ includes model $i$.

Let us now consider the posterior distribution function for some parameter $x$, common to all models in the set.
Within the $i$-th model, after applying the dataset $d$
the parameter $x$ has a posterior distribution $p(x|d,\mathcal{M}_i)$.
Using the posterior probability for each model, we can compute the model-marginalized
posterior distribution $p(x|d)$ for the parameter $x$ over the entire set of models by using:
\begin{equation}
\label{eq:mmposterior_def}
p(x|d)
\equiv
\sum_i
p(x|d,\mathcal{M}_i)
p_i
=
\frac{\sum_i
p(x|d,\mathcal{M}_i)
B_{i0}
}{
\sum_j B_{j0}
}
\,.
\end{equation}

Since one of the goals of our analysis is to study neutrino mass bounds,
we have to face the fact that the likelihoods under consideration (see next section)
are open with respect to $\log(\mnu)$, in the sense that
no lower limit on this quantity emerges from current cosmological probes.
In order to avoid the prior dependence of the credible intervals,
see e.g.~\cite{Heavens:2018adv,Stocker:2020nsx,Hergt:2021qlh,diValentino:2022njd},
one can adopt a method called \emph{relative belief updating ratio}
$\mathcal{R}$, see e.g.~\cite{Gariazzo:2019xhx}.
Given a parameter $x$ for which the likelihood is open,
it is convenient to define
\begin{equation}
\label{eq:R}
\mathcal{R}(x_1, x_2|d,\mathcal{M})
=
\frac{
p(x_1|d,\mathcal{M})/\pi(x_1|\mathcal{M})
}{
p(x_2|d,\mathcal{M})/\pi(x_2|\mathcal{M})
}\,,
\end{equation}
where $p(x_i|d,\mathcal{M})$ and $\pi(x_i|\mathcal{M})$ are respectively the values of the posterior
and prior distributions\footnote{We assume that the prior of $x$ is independent on the other parameters in the model.}
at $x_i$ within some model $\mathcal{M}$ and when considering the dataset $d$.
In case of simple posteriors and priors, it is easy to show that $\mathcal{R}$ is equivalent
to a likelihood ratio test, but for more complicated cases the results
can deviate from those one could obtain with the frequentist method,
that only considers the likelihood in two points, instead of the full parameter space volume.
At the computational level, however, $\mathcal{R}$ can be obtained from a Markov Chain Monte Carlo (MCMC)
without the need of dedicated log-likelihood minimizations.
Moreover, it is possible to show that $\mathcal{R}$ is equivalent to a Bayes factor between two submodels of model $\mathcal{M}$,
each of them obtained by fixing the value of $x$ to $x_1$ and $x_2$, respectively.

The above definition of $\mathcal{R}(x_1,x_2|d)$
can be easily extended to perform a model marginalization.
Assuming that the parameter $x$ is shared among all the models of the set,
and that its prior is the same in all models, one can write:
\begin{equation}
\label{eq:mmR}
\mathcal{R}(x_1, x_2|d)
=
\frac{
p(x_1|d)/\pi(x_1)
}{
p(x_2|d)/\pi(x_2)
}\,,
\end{equation}
where $p(x|d)$ is the model-marginalized posterior in Eq.~\eqref{eq:mmposterior_def}.
In our specific case, in order to study neutrino mass bounds we will consider $x_2=0$
and show the dependency $\mathcal{R}(\mnu, 0|d)$.

\section{Datasets}
\label{sec:data}

Our baseline data-sets consist of:

\begin{itemize}

\item The Atacama Cosmology Telescope DR4 likelihood~\cite{ACT:2020frw,ACT:2020gnv}, combined with WMAP 9-year observations data~\cite{Hinshaw:2012aka,WMAP:2012fli} and a Gaussian prior on $\tau = 0.065 \pm 0.015$, as done in Ref.~\cite{ACT:2020gnv}. This data-set is always considered in combination with the Pantheon catalog which includes a collection of 1048 B-band observations of the relative magnitudes of Type Ia supernovae~\cite{Pan-STARRS1:2017jku}, as well as together with Baryon Acoustic Oscillations (BAO) and Redshift Space Distortions (RSD) measurements obtained from a combination of the spectroscopic galaxy and quasar catalogs of the Sloan Digital Sky Survey (SDSS)~\cite{BOSS:2012dmf} and the more recent eBOSS DR16 data\footnote{It's worth noting that when we combine the DR12 data with the eBOSS DR16 data, we only use the first two redshift bins from DR12 in the $0.2 < z < 0.6$ range, which are further divided into the $0.2 < z < 0.5$ and $0.4 < z < 0.6$ regions.}~\cite{BOSS:2016wmc,eBOSS:2020yzd}. For simplicity, we refer to this combined data-set as 'ACT' in the following analysis.

\item The South Pole Telescope polarization measurements SPT-3G~\cite{SPT-3G:2021eoc} of the TE EE spectra are combined with WMAP 9-year observations data~\cite{Hinshaw:2012aka,WMAP:2012fli} and a Gaussian prior on $\tau = 0.065 \pm 0.015$. Similar to the ACT case, we also include the Pantheon catalog of Type Ia supernovae~\cite{Pan-STARRS1:2017jku}, along with the BAO and RSD measurements obtained from the same combination of SDSS and eBOSS DR16 measurements~\cite{BOSS:2012dmf,BOSS:2016wmc,eBOSS:2020yzd}. This combined data-set is referred to as 'SPT' in our analysis.

\end{itemize}

\section{Models}
\label{sec:models}
A key point in our analysis is to derive robust bounds on the neutrino mass from high multipole CMB experiments marginalizing over a plethora of possible background cosmologies. Therefore, along with the six $\Lambda$CDM parameters (the amplitude $A_s$ and the spectral index $n_s$ of scalar perturbations, the baryon $\Omega_b h^2$ and the cold dark matter $\Omega_c h^2$ energy densities, the angular size of the sound horizon at recombination $\theta_{\rm MC}$ and the reionization optical depth, $\tau$), we include the sum of neutrino masses $\sum m_\nu$ and, in a second step, we also add the number of relativistic degrees of freedom, parameterized via $\neff$.  We model $\sum m_{\nu}$ as three massive neutrinos with a normal hierarchy%
\footnote{Apart from the difference in the lower limit on the sum of neutrino masses, the assumption on the mass ordering is not expected to alter our results. The differences induced by considering degenerate mass eigenstates versus a more correct mass ordering are exceedingly small to be detected even by future experiments \cite{Archidiacono:2020dvx}, thus they cannot alter our conclusions based on current observations.}, but we assume an uninformative prior as in \autoref{tab.Priors}. When $\sum m_\nu<0.06$~eV, one massive and two massless neutrinos are considered.
 \neff\ is parametrized as the active neutrinos contribution to the energy density when
they are relativistic. The expected value is $\neff = 3.044$ \cite{Akita:2020szl,Froustey:2020mcq,Bennett:2020zkv} (see also \cite{Cielo:2023bqp}), and any additional contribution will be given by extra dark radiation coming from additional degrees of freedom
at recombination, such as relativistic dark matter particles or GWs. It is possible
to have $\neff < 3.044$ for 3 active massless neutrinos in case of low-temperature re-
heating \cite{deSalas:2015glj}, so we do not impose a lower prior
on this parameter.
We then explore a number of possible extensions of these close-to-minimal neutrino cosmologies, enlarging the parameter space including one or more parameters, such as
a running of the scalar index ($\alpha_s$),
a curvature component ($\Omega_k$), a non-vanishing tensor-to-scalar ratio ($r$), the dark energy equation of state parameters ($w_0$ and $w_a$), the lensing amplitude ($A_{\rm lens}$), the primordial helium fraction ($Y_{He}$) and the effective sterile neutrino mass ($m_{\nu,s}^{\rm eff}$) (see \autoref{tab.Priors} for the priors adopted in the cosmological parameters). 
A word of caution is mandatory here. Notice that the models here considered assume that neutrinos interact exclusively via weak interacting processes, excluding, for simplicity a vast number of very appealing non-standard neutrino cosmologies in which neutrinos exhibit interactions beyond the Standard Model (SM) of elementary particles. Should that be the case, the cosmological neutrino mass bounds will be considerably relaxed, see Ref.~\cite{Alvey:2021xmq} for a a complete review of possible scenarios.  Examples of beyond the SM neutrino interacting cosmologies include possible decays or annihilations of neutrinos into lightest degrees of freedom~\cite{Beacom:2004yd,Serpico:2007pt} in which a  significant relaxation of the neutrino mass constraint is found ($\sum m_\nu< 0.42$~eV at $95\%$~CL)~\cite{FrancoAbellan:2021hdb}, late-time neutrino mass generation models~\cite{Chacko:2004cz,Dvali:2016uhn}, where cosmological constraints could be completely evaded~\cite{Huang:2022wmz}, and long-range neutrino interactions~\cite{Esteban:2021ozz,Esteban:2022rjk}. Keeping the former restrictions in mind, in the following, we describe the possible extensions considered here, all of them assuming that neutrinos do not show interactions beyond the SM and therefore restricting ourselves to the simplest subset of possible fiducial cosmologies:

\begin{itemize}
\item Curvature density, $\Omega_k$. Recent data analyses of the CMB temperature and polarization spectra from Planck 2018 team exploiting the baseline \emph{Plik} likelihood suggest that our Universe could have a closed geometry at more than three standard deviations~\cite{Planck:2018vyg,Handley:2019tkm,DiValentino:2019qzk,Semenaite:2022unt}. These hints mostly arise from TT observations, that would otherwise show a lensing excess~\cite{DiValentino:2020hov,Calabrese:2008rt,DiValentino:2019dzu}. 
    In addition, analyses exploiting the \emph{CamSpec} TT likelihood~\cite{Efstathiou:2019mdh,Rosenberg:2022sdy} point to a closed geometry of the Universe with a significance above 99\% CL. Furthermore, an indication for a closed universe is also present in the BAO data, using Effective Field Theories of Large Scale Structure~\cite{Glanville:2022xes}. These recent findings strongly motivate to leave the curvature of the Universe as a free parameter~\cite{Anselmi:2022uvj} and obtain limits on the neutrino mass and abundances in this context. 

    \item The running of scalar spectral index, $\alpha_s$. In simple inflationary models, the running of the spectral index is a second order perturbation and it is typically very small. However, specific models can produce a large running over a range of scales accessible to CMB experiments. Indeed, a non-zero value of $\alpha_s$ alleviates the $\sim 2.7\sigma$ discrepancy in the value of the scalar spectral index $n_{\rm s}$ measured by \emph{Planck} ($n_{\rm s}=0.9649\pm 0.0044$)~\cite{Planck:2018vyg,Forconi:2021que} and by the \emph{Atacama Cosmology Telescope}  (ACT) ($n_{\rm s}=1.008\pm 0.015$)~\cite{ACT:2020gnv}, see Refs.~\cite{DiValentino:2022rdg, DiValentino:2022oon,Giare:2022rvg}. As previously stated, the different fiducial cosmologies considered here are the most economical ad simplest scenarios to be address, enough to illustrate the main goal of the manuscript. We have therefore not considered here models in which the primordial power spectrum is further modified not only with a running of the scalar spectral index but also either via features in its shape or by a description via a number of nodes in $k$~\cite{Canac:2016smv,DiValentino:2016ikp}, addition that will be performed elsewhere.

\item The tensor-to-scalar ratio $r$. Within  this extended model, we allow the tensor
perturbations to vary as well, along with scalar ones. Contributions to $r$ arise from the CMB B-mode polarization from either primordial gravitational waves or gravitational weak lensing. We therefore expect a larger effect for the observational data set where the polarization input is more relevant (as it is the case of SPT). 
    
    \item Dynamical Dark Energy equation of state. 
    Cosmological neutrino  mass bounds become weaker if the dark energy equation of state is taken as a free parameter. Even if current data fits well with the assumption of a cosmological constant within the minimal $\Lambda$CDM scenario, the question of having an equation of state parameter different from $ -1 $ remains certainly open. Along with constant dark energy equation of state models, in this manuscript we shall also consider the possibility of having a time-varying $ w(a) $ described by the Chevalier-Polarski-Linder parametrizazion (CPL)~\cite{Chevallier:2000qy,Linder:2002et}:
    \begin{equation}
    \label{eq:cpl}
     	w(a) = w_0 + (1-a)w_a\,,
    \end{equation}
    where $ a $ is the scale factor, equal to $ a_0 = 1 $ at the present time, $ w(a_0)=w_0 $ is the value of the equation of state parameter today. Dark energy changes the distance to the CMB consequently pushing it further (closer) if $w < -1$ ($w > -1$) from us. This effect can be balanced by having a larger matter density or, equivalently, by having more massive hot relics, leading to less stringent bounds on the neutrino masses. Accordingly, the mass bounds of cosmological neutrinos will  become weaker if the dark energy equation of state is taken as a free parameter. 

    \item The lensing amplitude $A_{\rm lens}$. CMB anisotropies get blurred due to gravitational lensing by the large scale structure of the Universe: photons from different directions are mixed and the peaks at large multipoles are smoothed. The amount of lensing is a precise prediction of the $\Lambda$CDM model: the consistency of the model can be checked by artificially increasing lensing by a factor $A_{\rm{lens}}$~\cite{Calabrese:2008rt} (\emph{a priori} an unphysical parameter). Within the $\Lambda$CDM picture, $A_{\rm{lens}}=1$. Planck CMB data shows a preference for additional lensing. Indeed, the reference analysis of temperature and polarization anisotropies suggest $A_\mathrm{lens} > 1$ at 3$\sigma$.  The lensing anomaly is robust against changes in the foreground modeling in the baseline likelihood, and was already discussed in previous data releases, although it is currently more significant due to the lower reionization optical depth preferred by the Planck 2018 data release. A recent result from the Atacama Cosmology Telescope is compatible with $A_\mathrm{lens}=1$~\cite{ACT:2020gnv}, but the results are consistent with Planck within uncertainties. Barring systematic errors or a rare statistical fluctuation,  the lensing anomaly could have an explanation within new physics scenarios.  Closed cosmologies~\cite{DiValentino:2019qzk} have been shown to solve the internal tensions in Planck concerning the cosmological parameter values at different angular scales, alleviating the $A_{\rm{lens}}$ anomaly. Neutrinos strongly affect CMB lensing and therefore their mass is degenerate with its amplitude, showing $\sum m_\nu$ and  $A_{\rm lens}$ a positive correlation, since increasing the neutrino mass reduces the smearing of the acoustic peaks, as a larger value of $\sum m_\nu$ increases the suppression to the small scale matter power~\cite{RoyChoudhury:2019hls,Sgier:2021bzf,RoyChoudhury:2018vnm,DiValentino:2021imh,DiValentino:2019dzu}.
    Also, non-standard long-range neutrino properties can lead to unexpected lensing and dilute the preference for $A_\mathrm{lens} \neq 1$~\cite{Esteban:2022rjk}.

    \item The helium fraction $Y_{\rm He}$. It is very well-known that the number if relativistic degrees of freedom is degenerate with the primordial helium fraction, due to the effect of these two parameters in the CMB damping tail. Silk
     damping refers to the suppression in power of the CMB temperature anisotropies on scales smaller than the photon diffusion length. 
    Varying both $N_{\rm {eff}}$ and the fraction of baryonic mass in Helium (that is, $Y_{\rm He}$) changes the ratio of Silk damping to sound horizon scales, leading to a degeneracy between these two parameters. Therefore, a much larger error on the neutrino abundance  (parameterized via $N_{\rm {eff}}$) is expected when considering $Y_{\rm He}$ a free parameter in the extended cosmological model.

    \item The effective sterile neutrino mass $m_{\nu,s}^{\rm eff}$. Finally, we should also consider the case in which the additional degrees of freedom refer to massive sterile neutrino states. If the extra massive sterile neutrino state has a thermal spectrum, its physical mass is $(\Delta N_{\rm{eff}})^{-3/4} m_{\nu,s}^{\rm eff}$, while in case of a non-resonant production~\cite{Dodelson:1993je} the physical mass is $(\Delta N_{\rm{eff}})^{-1} m_{\nu,s}^{\rm eff}$.  The constraints on the neutrino parameters will obviously depend on the amount of hot dark matter in the form of additional (sterile) massive neutrino states.
\end{itemize}

One may be worried that the considered selection of models is somewhat incomplete.
For example, one might expect that we presented an analysis with more than one or all the above-mentioned parameters varying at the same time, in addition to the base model.
In such high-dimensional models, we observe a natural reduction in the constraining power on the total neutrino mass (see e.g.~\cite{DiValentino:2019dzu}).
As a result, incorporating extensions with many parameters might,
in principle, lead to a weakening of the bound on the neutrino mass. However, it
is essential to take into account that high-dimensional models with numerous varying parameters are typically disfavored based on the Occam's razor principle. To
put it more quantitatively, models with fewer parameters generally present stronger
Bayesian evidence compared to those with a higher number of parameters. Consequently, in practice, when marginalizing over the model, scenarios with numerous
parameters carry significantly less weight in the average of the posteriors, and we do
not expect them to influence significantly the results derived in this study.

\section{Numerical implementation}
\label{sec:implementation}

To derive observational constraints within the different extended cosmological models discussed in the previous section, we perform Markov Chain Monte Carlo (MCMC) analyses using the publicly available package \texttt{CosmoMC}~\citep{Lewis:2002ah,Lewis:2013hha} and computing the theoretical model with the latest version of the Boltzmann code \texttt{CAMB}~\citep{Lewis:1999bs,Howlett:2012mh}. 

The prior distributions for all the parameters involved in our MCMC sampling are reported in \autoref{tab.Priors}. We choose uniform priors across the range of variation, except for the optical depth where a gaussian prior $\tau = 0.065 \pm 0.015$ is adopted, as indicated in \autoref{sec:data}. In addition, we consider two different cases for the DE equation of state $w_0$: in one case we assume the flat prior reported in \autoref{tab.Priors} while, in the other case, we force $w_0>-1$.

For model comparison, we compute the Bayesian Evidence of the different models and estimate the Bayes factors using the publicly available package \texttt{MCEvidence}.\footnote{\href{https://github.com/yabebalFantaye/MCEvidence}{github.com/yabebalFantaye/MCEvidence}~\cite{Heavens:2017hkr,Heavens:2017afc}}

In this regard, it is important to note that the estimation of Bayes factors, based on the MCMC results, is weakly dependent on the chosen priors for cosmological parameters. The impact of a uniform prior on $\sum m_{\nu}$ has been extensively discussed in the literature and we refer to Refs.~\cite{Simpson:2017qvj,Schwetz:2017fey,Gariazzo:2018pei,Gariazzo:2019xhx,Heavens:2018adv,GAMBITCosmologyWorkgroup:2020rmf,Hergt:2021qlh,Gariazzo:2022ahe} for further details (see also Ref.~\cite{Galloni:2022mok} for similar considerations on the tensor amplitude $r$). On the other hand, in a recent study~\cite{DiValentino:2022edq}, we have evaluated the difference in the Bayesian factors estimated using \texttt{MCEvidence} and those obtained by means of proper nested sampling algorithms such as \texttt{PolyChord}~\cite{Handley:2015fda,Handley:2015aa}. In particular, we employed a dedicated set of simulations where a 3D multi-modal Gaussian likelihood was used to constrain a 3-parameter model as the simplest case, and then compared it with two different models featuring 4 parameters. The Bayesian evidences obtained with \texttt{MCEvidence} were found to be consistently larger than those obtained with \texttt{PolyChord} by a factor of approximately $e$. When calculating the logarithm of the Bayes factors, the discrepancy between \texttt{MCEvidence} and \texttt{PolyChord} ranged between $-0.5$ and $0.2$ for all the considered cases. We consider this difference not significant enough to have a substantial impact on the marginalized bounds derived in this work.

\begin{table}[t]
	\begin{center}
		\renewcommand{\arraystretch}{1.5}
		\begin{tabular}{l@{\hspace{0. cm}}@{\hspace{3.5 cm}} c}
			\hline
			\textbf{Parameter}    & \textbf{Prior} \\
			\hline\hline
			$\Omega_{\rm b} h^2$         & $[0.005\,,\,0.1]$ \\
			$\Omega_{\rm c} h^2$         & $[0.001\,,\,0.99]$ \\
			$100\,\theta_{\rm {MC}}$     & $[0.5\,,\,10]$ \\
                $\tau$                       & $0.065 \pm 0.015$ \\
			$\log(10^{10}A_{\rm S})$     & $[1.6\,,\,3.9]$ \\
			$n_{\rm s}$                  & $[0.8\,,\, 1.2]$ \\
                \hline
   			$\sum m_{\nu}$ [eV]          & $[0\,,\,5]$\\
                $N_{\rm eff}$     	         & $[0.05\,,\,10]$\\
                \hline
                $r$     	                 & $[0\,,\,3]$\\
			$\Omega_{\rm k} $     	     & $[-0.3\,,\,0.3]$\\
			$w_0$                        & $[-3\,,\,1]$ \\
			$w_a$                        & $[-3\,,\,2]$ \\
                $\alpha_{\rm s}$             & $[-1\,,\, 1]$ \\
                $A_{\rm lens}$               & $[0\,,\, 10]$ \\
                $Y_{\rm He}$                 & $[0.1\,,\,0.5]$ \\
                $m_{\nu,s}^{\rm eff}$        & $[0\,,\,3]$ \\
			\hline\hline
		\end{tabular}
		\caption{List of uniform prior distributions for cosmological parameters.}
		\label{tab.Priors}
	\end{center}
\end{table}

\section{Results}
\label{sec:results}


\subsection{Total neutrino mass}

\begin{table*}
\centering
\renewcommand{\arraystretch}{1.5}
\resizebox{0.95\textwidth}{!}{\begin{tabular}{l @{\hspace{0.4 cm}} lccccccccc}
\toprule
\textbf{Cosmological model} & &\boldmath{$-\ln\rm{BF}$} & \boldmath{$\sum m_{\nu}$\textbf{[eV]}} & \boldmath{$N_{\text{eff}}$} & \boldmath{$\Omega_{k}$} &  \boldmath{$\alpha_s$} &  \boldmath{$r$} &\boldmath{$w_0$}& \boldmath{$w_a$}& \boldmath{$A_{\text{lens}}$} \\
\hline\hline

$+\sum m_{\nu}$ & ACT  &$5.10$   &  $< 0.176$ & -- & -- &-- &-- &--&-- &--\\
&SPT &$5.96$   &  $< 0.197$ & -- &-- &-- & -- &-- &-- &--\\[2ex]

$+\sum m_{\nu} + N_{\rm eff}$& ACT  &$2.59$   &  $< 0.155$ & $2.78\pm 0.25$ & -- &-- &-- &--&-- &--\\
&SPT &$2.96$   &  $< 0.238$ & $3.20\pm 0.31$ &-- &-- & -- &-- &-- &--\\[2ex]

$+\sum m_{\nu} + \Omega_k$& ACT  &$0.56$   & $< 0.271$ & -- & $0.0027\pm 0.0032$ &-- &-- &--&-- &--\\
&SPT &$1.08$   & $< 0.264$ & -- &$0.0013^{+0.0030}_{-0.0034}$ &-- & -- &-- &-- &--\\[2ex]

$+\sum m_{\nu} + \alpha_s$& ACT  &$1.16$   & $< 0.196$ & -- & -- &$0.0117\pm 0.0076$ &-- &--&-- &--\\
&SPT &$1.06$   & $< 0.215$ & -- &-- &$0.0054\pm 0.0092$ & -- &-- &-- &--\\[2ex]

$+\sum m_{\nu} + r$& ACT  &$1.60$   & $<0.189$ & -- & -- &-- & $<0.184$ &--&-- &--\\
&SPT &$2.55$   & $< 0.217$ & -- &-- &-- & $<0.207$ &-- &-- &--\\[2ex]

$+\sum m_{\nu} + w_0$& ACT  &$1.49$   & $< 0.244$ & -- & -- &-- &-- &$ -1.036\pm 0.037$&-- &--\\
&SPT &$1.74$   & $< 0.259$ & -- &-- &-- & -- & $-1.024\pm 0.037$ &-- &--\\[2ex]

$+\sum m_{\nu} + (w_0>-1)$& ACT  &$0.33$   & $< 0.159$ & -- & -- &-- &-- & $< -0.951$ &-- &--\\
&SPT &$1.31$   & $< 0.180$ & -- &-- &-- & -- & $< -0.946$ &-- &--\\[2ex]

$+\sum m_{\nu} + w_0 + w_a$& ACT  &$0.0$   & $< 0.361$ & -- & -- &-- &-- & $-0.951^{+0.082}_{-0.096}$ & $-0.46^{+0.52}_{-0.35}$ &--\\
&SPT &$0.0$   & $< 0.353$ & -- &-- &-- & -- & $-0.963^{+0.080}_{-0.094}$ & $-0.34^{+0.50}_{-0.33}$ &--\\[2ex]

$+\sum m_{\nu} + A_{\rm lens}$& ACT  &$1.04$   & $< 0.184$ & -- & -- &-- &-- &--&-- &$0.997^{+0.077}_{-0.086}$\\
&SPT &$2.14$   & $ < 0.176$ & -- &-- &-- & -- &-- &-- &$0.916\pm 0.074$\\[2ex]

\hline

marginalized & ACT  &--   & $< 0.24$ & -- & -- &-- &-- &--&-- &--\\
&SPT &--   & $ < 0.30$ & -- &-- &-- & -- &-- &-- &--\\[2ex]

\hline
\bottomrule
\end{tabular}}
\caption{Constraints on \mnu\ and other parameters from the $\Lambda$CDM + $\sum m_{\nu}$ model and its extension, and negative logarithms of the Bayes factors with respect to the preferred model within each data combination. Upper limits are at 95\% CL, while two-sided ones are at 68\% CL.}
\label{tab:mnu_res}
\end{table*}

\begin{figure}
\centering
\includegraphics[width=\columnwidth]{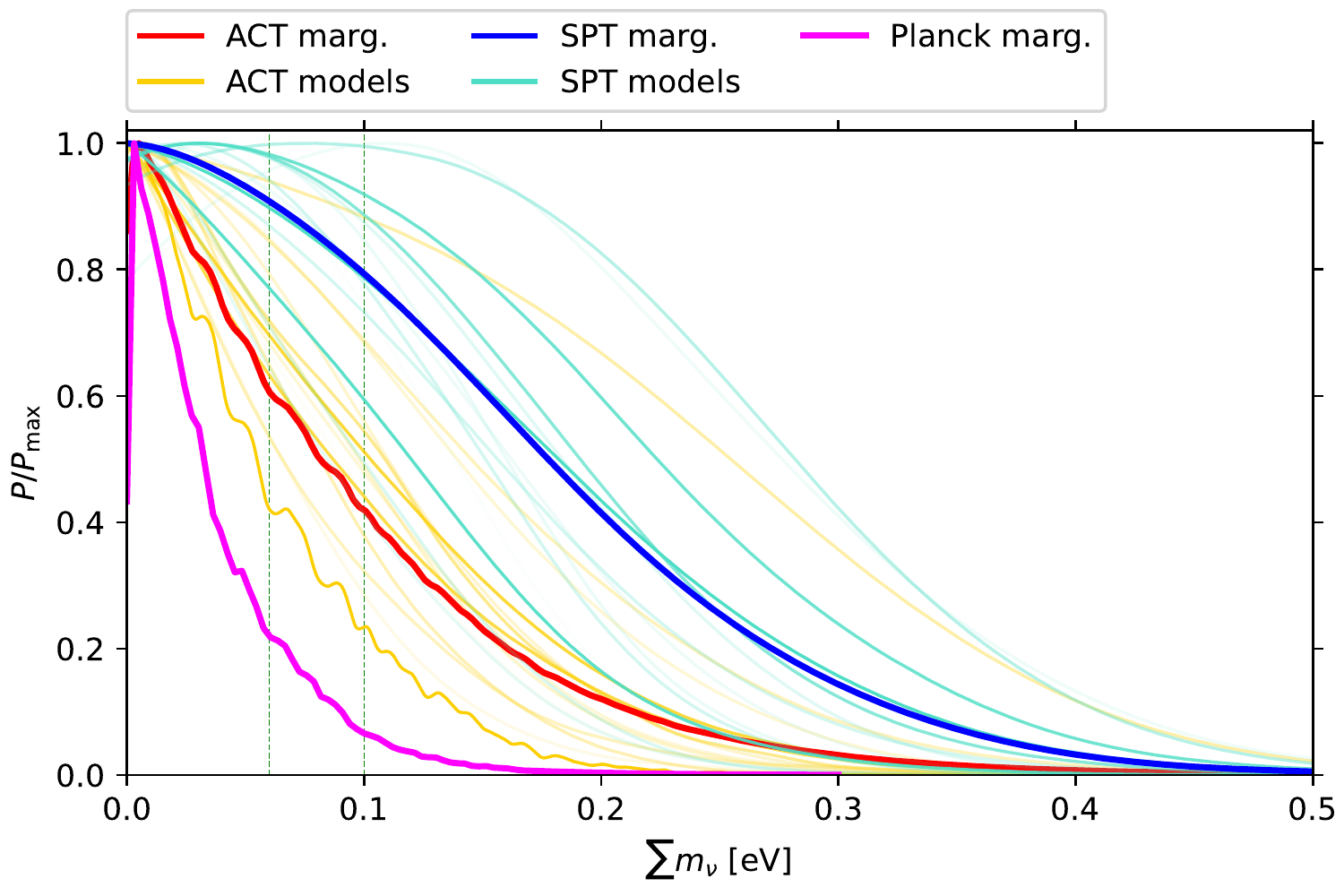}
\includegraphics[width=\columnwidth]{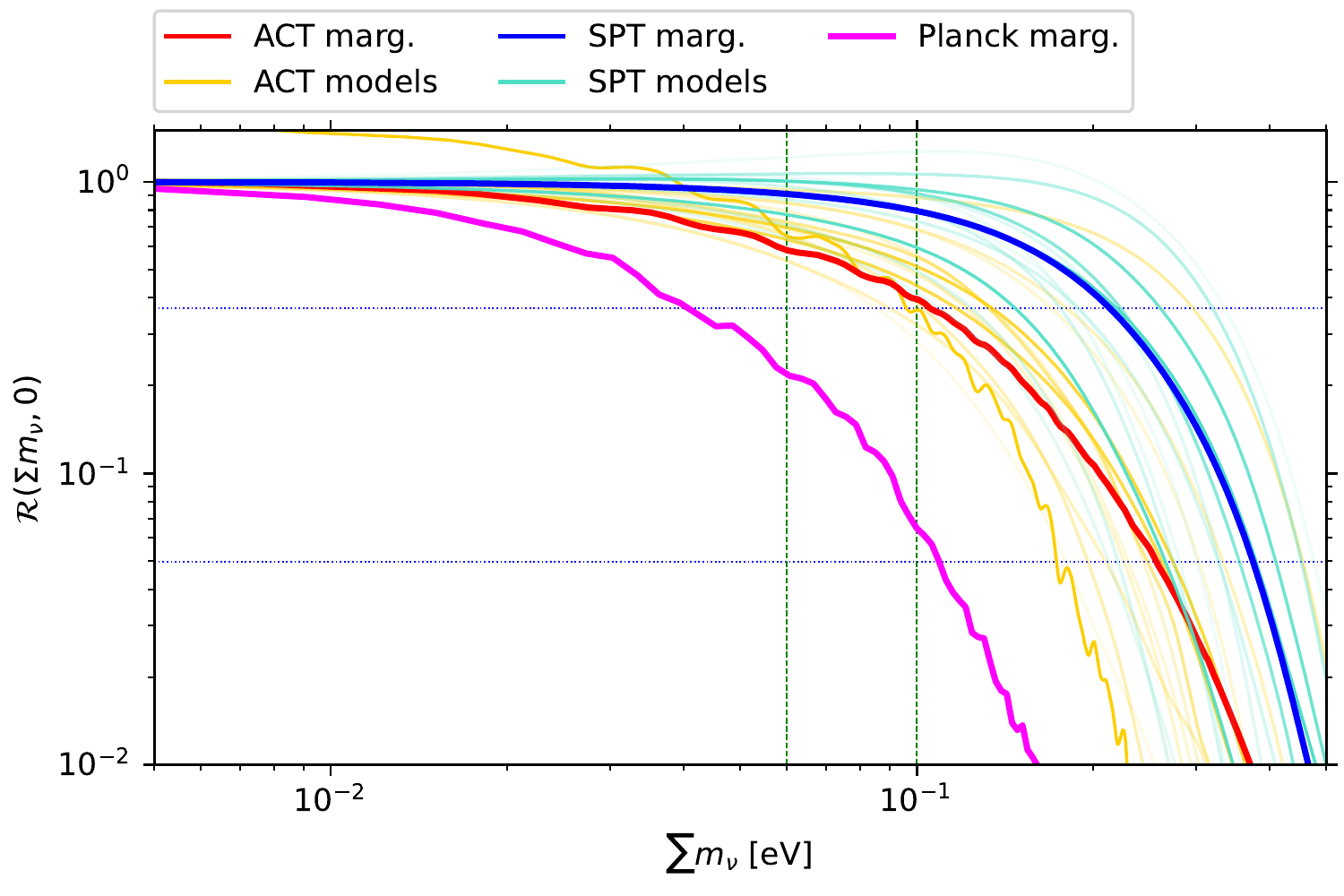}
\caption{
\label{fig:mnu_1d}
\textit{Upper panel:} Posteriors of \mnu\ within each model for the ACT (yellow) and SPT (light blue) data combinations,
plus model-marginalized posteriors for the ACT (red), SPT (blue) and Planck data combination (magenta, from~\cite{diValentino:2022njd}).
The vertical dotted lines indicate the minimum values for \mnu\ obtained within normal (0.06~eV) or inverted (0.1~eV) neutrino mass ordering.
\textit{Lower panel:}
Same color coding than in the upper panel, but in this case the quantity depicted is $\mathcal{R}(\mnu,0)$.
The horizontal lines show values at which $\mathcal{R}$ is either $e^{-1}$ or $e^{-3}$.
}
\end{figure}

\autoref{tab:mnu_res} summarizes the 95\% CL limits on the total neutrino mass $\sum m_{\nu}$ obtained for the two data-sets considered in this study, within the different cosmological models discussed in \autoref{sec:models}. We also present the 68\% CL constraints on the additional free parameters within each fiducial cosmology. 

Comparing the constraints obtained for different parameters, we observe that the constraining power of the two data-sets is quite similar, although ACT is more restrictive than SPT when dealing with the total neutrino mass. Indeed, within a minimal $\Lambda$CDM+$\sum m_{\nu}$ extension, at 95\% CL we get $\sum m_{\nu}<0.176$ eV from ACT and $\sum m_{\nu}<0.197$ eV from SPT, respectively. Notice that these limits, even if  they are less competitive than those found with Planck data, are still very constraining. 

Regarding the results for the additional parameters, it is worth noting that both ACT and SPT are in agreement, within one standard deviation, with a flat spatial geometry ($\Omega_k=0$), a lensing amplitude consistent with its $\Lambda$CDM value ($A_{\rm{lens}}=1$), and a constant dark energy equation of state ($w_a=0$), matching also the expected value for a cosmological constant ($w_0=-1$). Nonetheless, ACT shows a decrease at slightly more than $1\sigma$ in the value of the effective number of relativistic degrees of freedom ($N_{\rm{eff}}=2.78\pm0.25$), while SPT is in good agreement with the expected value for this parameter. Concerning the inflationary sector of the theory, it is worth mentioning that both data-sets suggest a small and positive running of the spectral index ($\alpha_s$). However, while SPT is consistent with $\alpha_s=0$ within one standard deviation, ACT prefers a non-vanishing running ($\alpha_s=0.0117\pm0.0076$) at about $1.5\sigma$. Finally, regarding the amplitude of tensor modes, we obtain the 95\% CL limits of $r<0.184$ for ACT and $r<0.207$ for SPT. While ACT is more constraining than SPT on primordial tensor modes, these bounds are not very competitive when compared to the most recent updated limits ($r<0.035$ at 95\% CL from the joint analysis of BK18~\cite{BICEP:2021xfz,BICEPKeck:2022mhb} and Planck 2018 data), as expected, due to the absence of B-mode polarization in the data combinations considered here.

By following the methodology detailed in \autoref{sec:bayesian}, one can marginalize over this range of models and obtain a model-marginalized limit on the total neutrino mass for both data-sets. In the case of ACT, we find $\sum m_{\nu}<0.24$ eV at a 95\% CL, while for SPT, the limit is $\sum m_{\nu}<0.30$ eV at a 95\% CL, confirming that ACT provides stronger constraints. These results are depicted in \autoref{fig:mnu_1d}, where we show the marginalized posterior distribution function of the total mass of neutrinos for both experiments, along with the corresponding result obtained using Planck data.
Notice that one of the yellow lines is above $\mathcal{R}=1$ at small \mnu, because of an extremely mild preference for $\mnu\simeq0.007$~eV in the $+\mnu+\neff+w$ scenario when using ACT data.
The figure demonstrates that all experiments provide very competitive bounds, with ACT being more constraining than SPT but less constraining than Planck.\footnote{The purple line that reports the model-marginalized Planck results is taken from \cite{diValentino:2022njd}.} However, the most important feature to notice here is that \emph{each of them is robust, as they provide model-independent constraint, and they are consistent among themselves, clearly stating the robustness of current cosmological neutrino mass bounds.} 


\subsection{Effective number of relativistic neutrinos}

\begin{table*}
\centering
\renewcommand{\arraystretch}{1.5}
\resizebox{0.95\textwidth}{!}{\begin{tabular}{l @{\hspace{0.4 cm}} lccccccccc}
\toprule
\textbf{Cosmological model} & &\boldmath{$-\ln\rm{BF}$} & \boldmath{$N_{\text{eff}}$} & \boldmath{$\sum m_{\nu}$} & \boldmath{$\Omega_{k}$} &  \boldmath{$\alpha_s$} &  \boldmath{$m_{\nu,s}^{\rm eff}$} &\boldmath{$w_0$}& \boldmath{$w_a$}& \boldmath{$Y_{He}$} \\
\hline\hline
$+ N_{\rm eff}$& ACT  &$4.90$   & $2.77\pm 0.24$ & -- & -- &-- &-- &--&-- &--\\
&SPT &$6.44$   & $3.14\pm 0.29$ & -- &-- &-- & -- &-- &-- &--\\[2ex]

$+ N_{\rm eff}+\sum m_{\nu}$& ACT  &$1.18$   & $2.78\pm 0.25$ & $< 0.155$ & -- &-- &-- &--&-- &--\\
&SPT &$2.66$   & $3.20\pm 0.31$ & $< 0.238$ &-- &-- & -- &-- &-- &--\\[2ex]

$+ N_{\rm eff}+\Omega_k$& ACT  &$0.61$   & $2.68\pm 0.25$ & -- & $0.0034\pm 0.0029$ &-- &-- &--&-- &--\\
&SPT &$1.52$   & $3.13\pm 0.30$ & -- & $0.0002\pm 0.0029$ &-- & -- &-- &-- &--\\[2ex]

$+ N_{\rm eff}+\alpha_s$& ACT  &$0.80$   & $2.97\pm 0.29$ & -- & -- &$0.0100\pm 0.0091$ &-- &--&-- &--\\
&SPT &$1.98$   & $3.24\pm 0.33$ & -- &-- &$0.0076\pm 0.0099$ & -- &-- &-- &--\\[2ex]

$+ N_{\rm eff}+w_0$& ACT  &$1.63$   & $2.67\pm 0.25$ & -- & -- &-- &-- &$-1.047\pm 0.037$&-- &--\\
&SPT &$2.22$   & $ 3.12\pm 0.32$ & -- &-- &-- & -- &$-1.012\pm 0.036$&-- &--\\[2ex]

$+ N_{\rm eff}+(w_0>-1)$& ACT  &$0.0$   & $2.82\pm 0.23$ & -- & -- &-- &--&$<-0.955$&-- &--\\
&SPT &$2.04$   & $3.22\pm 0.30$ & -- &-- &-- & -- &$< -0.937$&-- &--\\[2ex]

$+ N_{\rm eff}+w_0+w_a$& ACT  &$0.38$   & $2.57\pm 0.26$ & -- & -- &-- &-- &$ -0.953\pm 0.088$& $-0.48^{+0.44}_{-0.35}$ &--\\
&SPT &$0.0$   & $3.06^{+0.31}_{-0.35}$ & -- &-- &-- & -- & $-0.987^{+0.076}_{-0.086}$ & $-0.14^{+0.40}_{-0.29}$ &--\\[2ex]

$+ N_{\rm eff}+Y_{He}$& ACT  &$3.05$   & $2.94^{+0.38}_{-0.45}$ & -- & -- &-- &-- &--&-- &$0.231^{+0.027}_{-0.024}$\\
&SPT &$6.24$   & $ 3.83^{+0.46}_{-0.56}$ & -- &-- &-- & -- &-- &-- &$0.187\pm 0.032$\\[2ex]

\hline
$+ N_{\rm eff}+m_{\nu,s}^{\rm eff}$& ACT  &ignored   & $< 3.37$ & -- & -- &-- &$< 0.556$&--&-- &--\\
&SPT &ignored   & $< 3.82$ & -- &-- &-- & $< 0.251$ &-- &-- &--\\[2ex]

\hline
marginalized & ACT  &--   & $2.79_{-0.28}^{+0.30}$ & -- & -- &-- &-- &--&-- &--\\
&SPT &--   & $3.21_{-0.33}^{+0.34}$ & -- &-- &-- & -- &-- &-- &--\\[2ex]
\hline
\bottomrule
\end{tabular}}
\caption{Same as in \autoref{tab:mnu_res} but for the $\Lambda$CDM + \neff\ model and its extensions. Notice that the model including $m_{\nu,s}^{\rm eff}$ as a free parameter is not included in the marginalization.}
\label{tab:nnu_res}
\end{table*}

\begin{figure}
\centering
\includegraphics[width=\columnwidth]{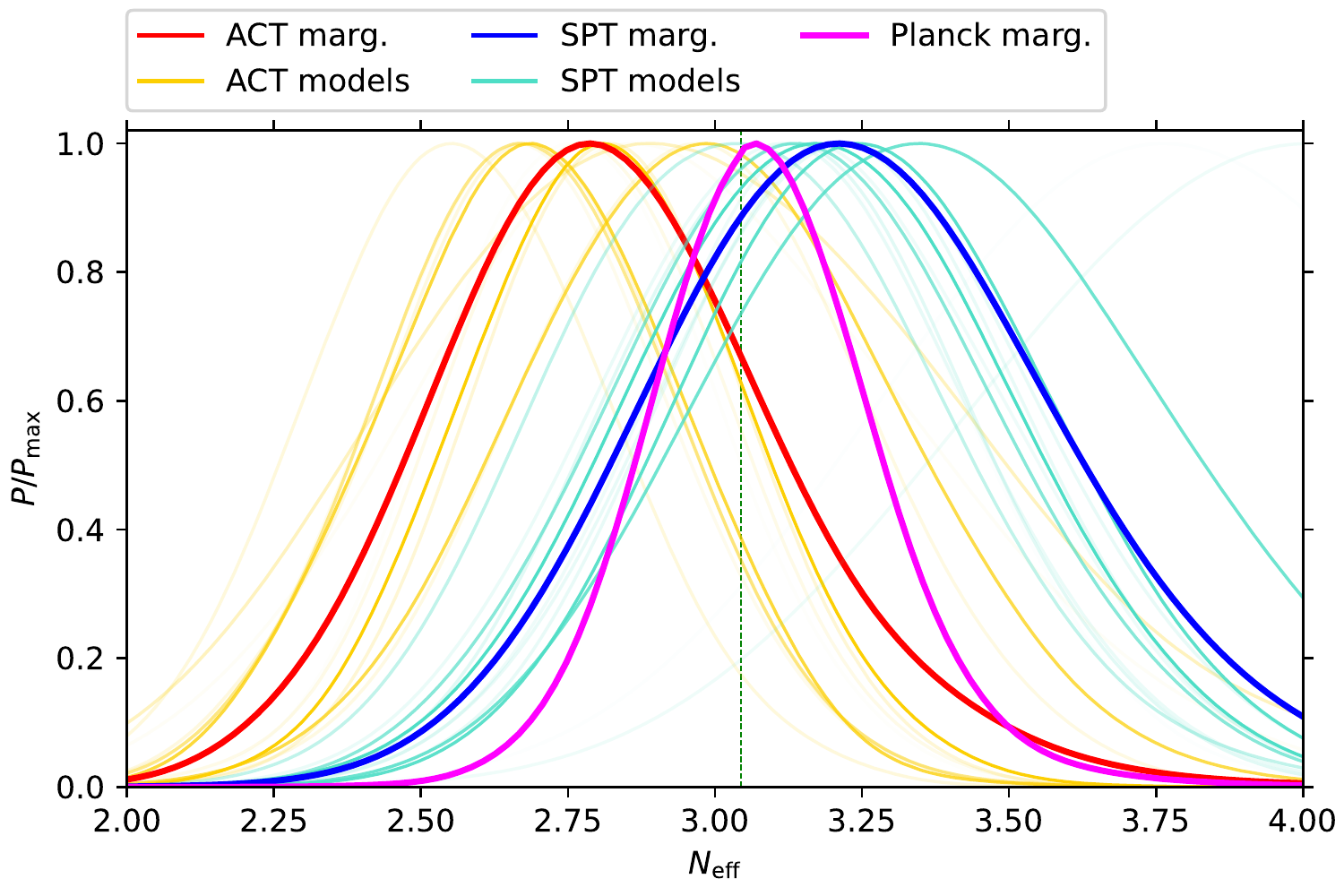}
\caption{
\label{fig:nnu_1d}
The same as in the upper panel of \autoref{fig:mnu_1d}, but for \neff.
The vertical line indicates $\neff=3.044$~\cite{Akita:2020szl,Froustey:2020mcq,Bennett:2020zkv}.
}
\end{figure}

\autoref{tab:nnu_res} presents the 68\% CL constraints on the effective number of relativistic degrees of freedom in the early universe, $N_{\rm{eff}}$. These constraints are derived under different background cosmologies, whose additional free parameters are also provided in the same table for both data-sets.

Firstly, it is worth noting that the results for $N_{\rm{eff}}$ obtained from the SPT data remain pretty consistent with the expected value from the standard model, and the constraints in the minimal $\Lambda$CDM+$N_{\rm{eff}}$ extension reads $N_{\rm{eff}}=3.14\pm0.29$ at 68\% CL. On the other hand, the agreement between the value predicted by ACT for $N_{\rm{eff}}$ and the standard model reference value depends, to some extent, on the background cosmology. Specifically, both in the minimal extension ($N_{\rm{eff}}=2.77\pm0.24$ at 68\% CL) and in more complicated cosmologies, we systematically observe the same mild shift of $N_{\rm{eff}}$ towards lower values compared to the standard model expectations, with a statistic significance ranging between one and two standard deviations.\footnote{ It is possible to have values of $N_{\rm{eff}}<3.044$ for three active massless neutrinos in case of low-temperature reheating (see Ref.~\cite{deSalas:2015glj}).}

Allowing $N_{\rm{eff}}$ to vary has implications for the constraints on the other parameters in the various cosmologies, sometimes changing the conclusions discussed in the previous subsection. For instance, while both ACT and SPT remain still consistent with a spatially flat Universe, when considering curvature, the preference of ACT for a smaller $N_{\rm{eff}}$ increases at the level of $1.5\sigma$. In addition, the ACT preference for a positive running is now significantly diluted. Most notably, we observe a preference at slightly more than one standard deviation for a phantom dark energy equation of state, as well as a $1\sigma$ indication for a dynamical behavior. Lastly, as previously stated, we have also considered an extension involving the effective sterile neutrino mass $m_{\nu,s}^{\rm eff}$. Since a sterile neutrino contributes to increase the effective number of relativistic degrees of freedom, in this case we necessarily have $N_{\rm eff}>3$. Therefore, we can only obtain only an upper bound on the additional contribution to the radiation energy-density in the early Universe, which is found to be $N_{\rm eff}<3.37$ and $N_{\rm eff}<3.82$ at $95\%$~CL for ACT and SPT, respectively.  The ACT preference for a lower value of $N_{\rm eff}$ is evident and translates into an upper limit on the mass of the sterile neutrino that is much less constraining for ACT ($m_{\nu,s}^{\rm eff}<0.556$ eV) than for SPT ($m_{\nu,s}^{\rm eff}<0.251$ eV), due to the strong degeneracy between these two parameters.

Similarly to what is done with the neutrino mass, also in this case we marginalize over the different models (excluding the extension involving the effective sterile neutrino which would artificially bias the results towards higher values of $N_{\rm eff}$) and obtain robust model-marginalized limits on $N_{\rm{eff}}$ for both data-sets. The results are summarized in \autoref{fig:nnu_1d}, which shows the posterior distribution function of $N_{\rm{eff}}$ for ACT and SPT, as well as their comparison with the result obtained by using the Planck data. From the figure, we can observe that ACT and SPT exhibit a similar precision, although ACT mildly favors $N_{\rm{eff}}<3$, resulting in a 68\% CL marginalized limit of $N_{\rm{eff}}=2.79_{-0.28}^{+0.30}$. On the other hand, SPT suggests $N_{\rm{eff}}>3$, yielding a marginalized bound of $N_{\rm{eff}}=3.21_{-0.33}^{+0.34}$ at 68\% CL. We would like to emphasize that for both data-sets, \emph{the marginalized limits show an excellent  agreement consistent with the standard model predictions}.


\subsection{Joint analysis of \boldmath{$\sum m_{\nu}$} and \boldmath{$N_{\rm{eff}}$} }

\begin{table*}
\centering
\renewcommand{\arraystretch}{1.5}
\resizebox{0.95\textwidth}{!}{\begin{tabular}{l @{\hspace{0.4 cm}} lccccccccc}
\toprule
\textbf{Cosmological model} & &\boldmath{$-\ln\rm{BF}$} & \boldmath{$\sum m_{\nu}$\textbf{[eV]}} & \boldmath{$N_{\text{eff}}$} & \boldmath{$\Omega_{k}$} &  \boldmath{$\alpha_s$} &  \boldmath{$m_{\nu,s}^{\rm eff}$} &\boldmath{$w_0$}& \boldmath{$Y_{He}$} \\
\hline\hline
$+\sum m_{\nu} + N_{\rm eff}$& ACT  &$4.71$   &  $< 0.155$ & $2.78\pm 0.25$ & -- &-- &-- &--&-- &--\\
&SPT &$4.74$   &  $< 0.238$ & $3.20\pm 0.31$ &-- &-- & -- &-- &-- &--\\[2ex]

$+\sum m_{\nu} + N_{\rm eff}+\Omega_k$& ACT  &$0.50$   & $< 0.230$ & $2.69\pm 0.25$ & $0.0038^{+0.0029}_{-0.0033}$ &-- &-- &--&-- \\
&SPT &$0.0$   & $< 0.309$ & $3.19\pm 0.32$ &$0.0012\pm 0.0034$&-- & -- &-- &-- \\[2ex]

$+\sum m_{\nu} + N_{\rm eff}+\alpha_s$& ACT  &$0.67$   & $< 0.224$ & $3.00\pm 0.32$ & -- & $0.0107\pm 0.0096$ &-- &--&-- \\
&SPT &$0.41$   & $< 0.325$ & $3.40^{+0.35}_{-0.42}$ &-- & $0.010\pm 0.011$ & -- &-- &-- \\[2ex]

$+\sum m_{\nu} + N_{\rm eff}+w_0$& ACT  &$1.61$   & $< 0.203$ & $2.68\pm 0.25$ & -- &-- &-- & $-1.050\pm 0.039$ & -- \\
&SPT &$0.79$   & $< 0.279$ & $3.16\pm 0.32$ &-- &-- & -- & $-1.021\pm 0.039$ &-- \\[2ex]

$+\sum m_{\nu} + N_{\rm eff}+(w_0>-1)$& ACT  &$0.0$   & $< 0.139$ & $2.82\pm 0.25$ & -- &-- &-- & $< -0.953$&-- \\
&SPT &$0.17$   & $< 0.217$ & $3.27\pm 0.32$ &-- &-- & -- & $< -0.937$ &-- \\[2ex]

$+\sum m_{\nu} + N_{\rm eff}+Y_{He}$& ACT  &$2.74$   & $< 0.188$ & $2.97^{+0.42}_{-0.52}$ & -- &-- &-- &--&$0.230^{+0.029}_{-0.024}$ \\
&SPT &$4.8$   & $< 0.365$ & $4.15^{+0.55}_{-0.69}$ &-- &-- & -- &-- &$0.179\pm 0.033$ \\[2ex]

\hline

$+\sum m_{\nu} + N_{\rm eff}+m_{\nu,s}^{\rm eff}$& ACT  &ignored   & $< 0.162$ & $< 3.38$ & -- &-- & $< 0.561$ &--&-- \\
&SPT & ignored   & $<0.224$ & $< 3.86$ &-- &-- & $< 0.232$ &-- &-- \\[2ex]
\hline

marginalized & ACT  &--   & $< 0.20$ & $2.79_{-0.28}^{+0.30}$ & -- &-- &-- &--&-- &--\\
&SPT &--   & $ < 0.30$ & $3.21_{-0.32}^{+0.35}$ &-- &-- & -- &-- &-- &--\\[2ex]

\hline
\bottomrule
\end{tabular}}
\caption{Same as in \autoref{tab:mnu_res} but for the $\Lambda$CDM + \mnu\ + \neff\ and its extensions. Notice that the model including $m_{\nu,s}^{\rm eff}$ as a free parameter is not included in the marginalization.}
\label{tab:mnu_nnu_res}
\end{table*}

\autoref{tab:mnu_nnu_res}  shows the results obtained after freely varying simultaneously both the total neutrino mass sum and the effective number of relativistic degrees of freedom in several possible fiducial cosmologies. 
Within the minimal $\Lambda$CDM+$\sum m_{\nu}$+$N_{\rm{eff}}$, ACT data provies a 95\% CL upper limit on $\sum m_{\nu} < 0.155$ eV and a 68\%CL constraint on $N_{\rm{eff}}=2.78\pm0.25$. Instead, for SPT observations, the bounds above are $\sum m_{\nu} < 0.238$ eV and $N_{\rm{eff}}=3.20\pm0.31$. These results confirm that ACT provides slightly stronger constraints compared to SPT, and show the very same mild ACT's preference for $N_{\rm{eff}}<3$ than that found in the previous section.

As concerns the constraints on the additional parameters, simultaneously varying $N_{\rm{eff}}$ and $\sum m_{\nu}$ does not change significantly the conclusions drawn in the previous sections: we find the same mild preference for positive values of the running of the spectral index and an indication at a slightly more than one standard deviation in favor of a phantom equation of state for dark energy. 

As in the previous sections, we marginalize here over the different fiducial cosmologies, finding joint model-marginalized limits for the two parameters. For ACT, the limits are $\sum m_{\nu}<0.2$~eV  and $N_{\rm{eff}}=2.79^{+0.30}_{-0.28}$, while for SPT they read $\sum m_{\nu}<0.3$ eV and $N_{\rm{eff}}=3.21^{+0.35}_{-0.32}$. These results are basically identical to those derived separately for the total neutrino mass and the effective number of relativistic degrees of freedom discussed in the previous subsections, clearly stating the important conclusion that \emph{current cosmological measurements are powerful enough to disentangle between the physical effects induced by the neutrino mass and those induced by the effective number of degrees of freedom, i.e.\ even if they are degenerate in terms of the total dark mass-energy density, the observables are sensitive to their (independent) footprints.}

\section{Conclusions}
\label{sec:conclusions}

As cosmological constraints on neutrino masses approach the lower bounds derived from neutrino oscillation data, the need for model-independent mass limits becomes increasingly important to evaluate the reliability of current cosmological measurements of the neutrino mass scale. In addition, it is crucial to assess the consistency between high-multipole and low-multipole Cosmic Microwave Background observations in measuring this scale to further evaluate their agreement and the global constraining power of the available data.

In this work, we consider a plethora of possible background cosmologies resulting from the inclusion of one or more parameters, such as
a running of the scalar index ($\alpha_s$),
a curvature component ($\Omega_k$), a non-vanishing tensor-to-scalar ratio ($r$), the dark energy equation of state parameters ($w_0$ and $w_a$), the lensing amplitude ($A_{\rm lens}$), the primordial helium fraction ($Y_{He}$) and the effective sterile neutrino mass ($m_{\nu,s}^{\rm eff}$). For each background cosmology, we derive up-to-date limits on neutrino masses and abundances by exploiting either the Data Release 4 of the Atacama Cosmology Telescope (ACT) or the South Pole Telescope polarization data from SPT-3G; always used in combination with WMAP 9-year observation data, the Pantheon catalog of Type Ia Supernovae and the BAO and RSD measurements obtained from the same combination of SDSS and eBOSS DR16 observations. Our methodology therefore provides  \emph{Planck-free} constraints on neutrino masses and abundances and serve as a calibration of Planck-independent data sets, addressing both the consistency and the robustness of current cosmological bounds on neutrino properties.

Following the Bayesian method outlined in \autoref{sec:bayesian}, we first marginalize over the different fiducial models  to determine a robust, model-marginalized constraint on the neutrino mass from  both ACT and SPT datasets. Our results are reported in \autoref{tab:mnu_res}. For the dataset involving the ACT measurements of the damping tail, we find $\sum m_{\nu}<0.24$ eV at 95\% CL, while when ACT is replaced with SPT the limit changes to $\sum m_{\nu}<0.30$ eV. In \autoref{fig:mnu_1d} we compare these results with those obtained by exploiting the Planck data in Ref.~\cite{diValentino:2022njd}. We notice that, while all experiments provide very competitive bounds, ACT is more constraining than SPT (but still less constraining than Planck).

We repeat the same procedure for the effective number of relativistic degrees of freedom $N_{\rm{eff}}$, obtaining the model-marginalized limits reported in \autoref{tab:nnu_res}. In this case we notice that ACT slightly favors $N_{\rm{eff}}<3$, resulting in a 68\% CL marginalized limit of $N_{\rm{eff}}=2.79_{-0.28}^{+0.30}$. On the other hand, SPT suggests $N_{\rm{eff}}>3$, yielding a marginalized bound of $N_{\rm{eff}}=3.21_{-0.33}^{+0.34}$ at 68\% CL. In both cases, however, the marginalized constraints are consistent within one standard deviation with the Standard Model predictions, as well as with the results obtained by using Planck (see also \autoref{fig:nnu_1d} where the marginalized 1D posteriors are shown for all cases).

Finally, we perform a joint analysis of both the total neutrino mass and the effective number of relativistic degrees of freedom, deriving model-marginalized limits when both parameters are allowed to vary in non-standard background cosmologies. In this case, our results are summarized in \autoref{tab:mnu_nnu_res} and read $\sum m_{\nu}<0.2$ eV, $N_{\rm{eff}}=2.79^{+0.30}_{-0.28}$ for ACT, and $\sum m_{\nu}<0.3$ eV, $N_{\rm{eff}}=3.21^{+0.35}_{-0.32}$ for SPT.

Our results reassess the robustness and reliability of current cosmological bounds on neutrino properties, at least  in the simplest extensions of the $\Lambda$CDM model. If neutrinos exhibit non-standard interactions beyond the canonical weak processes dictated by the SM of elementary particles, the limits could be significantly relaxed. Restricting ourselves to a plethora of the most economical scenarios, we find the neutrino mass and abundances bounds to be very stable, also when model--marginalizing over them. On the other hand, measurements of the CMB at different multipoles offer very similar and consistent results.

\begin{acknowledgments}
\noindent  
EDV is supported by a Royal Society Dorothy Hodgkin Research Fellowship. 
This article is based upon work from COST Action CA21136 Addressing observational tensions in cosmology with systematics and fundamental physics (CosmoVerse) supported by COST (European Cooperation in Science and Technology). 
This work has been partially supported by the MCIN/AEI/10.13039/501100011033 of Spain under grant PID2020-113644GB-I00, by the Generalitat Valenciana of Spain under grant PROMETEO/2019/083 and by the European Union's Framework Programme for Research and Innovation Horizon 2020 (2014–2020) under grant agreement 754496 (FELLINI) and 860881 (HIDDeN).
\end{acknowledgments}


\bibliography{biblio}	

\end{document}